\documentclass[inte,nonblindrev]{informs3} 

\OneAndAHalfSpacedXII 

\usepackage{natbib}
\bibpunct[, ]{(}{)}{,}{a}{}{,}%
\def\bibfont{\small}%

\graphicspath{{fig/}}
\usepackage{hyperref}
\usepackage{xcolor}
\usepackage{multirow}

\TheoremsNumberedThrough     

\EquationsNumberedThrough    

\begin{document}
	
	
	\RUNAUTHOR{Liu, Chen and Teo}
	
	\RUNTITLE{Limousine Service Management}
	
	\TITLE{Limousine Service Management: Capacity Planning with Predictive Analytics and Optimization}
	
	\ARTICLEAUTHORS{%
		\AUTHOR{Peng Liu}
		\AFF{Department of Statistics \& Applied Probability, National University of Singapore, \EMAIL{peng.liu@u.nus.edu}}
		\AUTHOR{Ying Chen}
		\AFF{Department of Mathematics, National University of Singapore, \EMAIL{matcheny@nus.edu.sg}
		\AUTHOR{Chung-Piaw Teo}
		\AFF{Institute of Operations Research and Analytics, National University of Singapore, \EMAIL{bizteocp@nus.edu.sg}}
}
	} 
	
	\ABSTRACT{%
		The limousine service in luxury hotels is an integral component of the whole customer journey in the hospitality industry. One of the largest hotels in Singapore manages a fleet of both in-house and outsourced vehicles around the clock, serving 9000 trips per month on average. The need for vehicles may scale up rapidly, especially during special events and festive periods in the country. The excess demand is met by having additional outsourced vehicles on standby, incurring millions of dollars of additional expenses per year for the hotel. Determining the required number of limousines by hour of the day is a challenging service capacity planning problem. In this paper, a recent transformational journey to manage this problem in the hotel is introduced, driving up to S\$3.2 million of savings per year with improved service level. The approach builds on widely available open-source statistical and spreadsheet optimization tools, along with robotic process automation, to optimize the schedule of its fleet of limousines and drivers, and to support decision-making for planners/controllers to drive sustained business value.
	}%
	
	
	
	\maketitle
	
	%
	\section{Introduction}\label{sec:intro}
	High-end hotels offer their guests a touch of chauffeur-driven luxury for direct airport and city transfers to major destinations within the city in a limousine vehicle of their choice. This service is often the first point of contact for the hotels to shape the service experience of their guests. Upon arrival, the VIPs are greeted by a driver and swiftly brought to their destination in comfort. Since most of the customers are hotel guests staying at suite level or above, a high level of vehicle availability is required to ensure a smooth customer journey. 
	Furthermore, before the start of a shift, drivers engaged for this service would need to ensure that the cars are  in good condition for the VIPs. Timely recording of each stage of a trip from start to end is also done to provide immediate feedback for the planning system to manage operational productivity and service availability.
	
\begin{figure}[htbp]
		\begin{center} 
			\includegraphics[width=14cm, height=8cm]{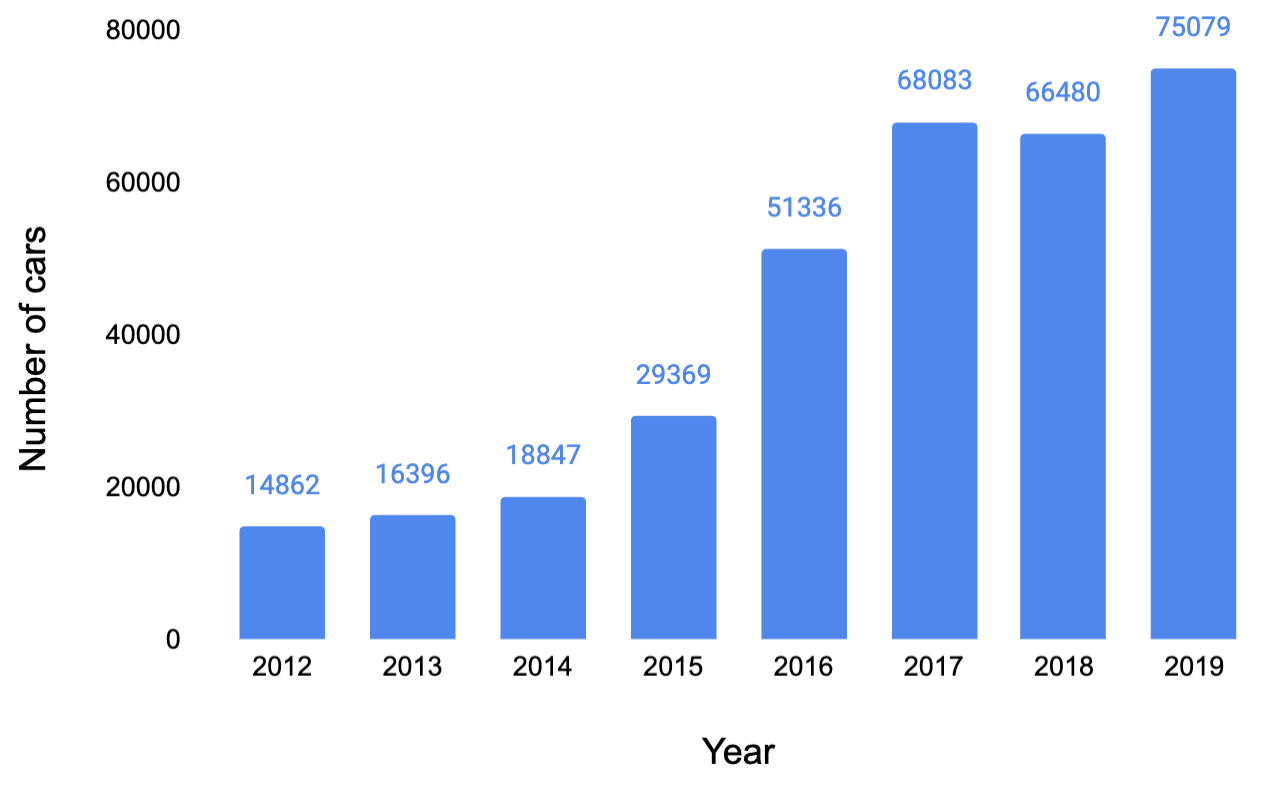}\hspace{0.1cm}
			\caption{The number of private hire cars in Singapore from 2012 to 2019, both self-drive and chauffeured. Source is \url{https://www.statista.com/statistics/953848/number-of-private-hire-cars-singapore/}}
			\label{fig:private} 
		\end{center} 
	\end{figure}

	Given stringent service standards, a new driver is usually trained for weeks and scored based on audit performance before being deployed to serve the VIPs. Outsourcing this service to popular ride hailing platforms is not viable, despite the fast growth of the latter in recent years (See 
	Figure \ref{fig:private} for the growth in the private car hire sector in Singapore, over the last eight years). Ride hailing platforms offer higher service availability due to the larger fleet size, but falls short of the exclusive service standards and experiences offered by professional, in-house drivers of the hotel.  
	
	This paper describes the transformation journey of a hotel in Singapore in managing this limousine service provision problem. Boasting a luxury clientele with high expectations, the hotel in our study is the leading destination for business, leisure and entertainment in Singapore. It offers an exquisite hotel, state-of-the-art convention and exhibition facilities, theatres, world-class entertainment and arguably the best shopping and dining experience in the region.  As Singapore's largest hotel with a significant volume of VIP clientele, the hotel constantly faces the challenge of managing its pool of limousines to ensure that services are available upon request, while ensuring operations are efficient and productive.  
	
	With more than 100,000 trips to serve annually, the limousine service provisioning problem for this hotel is a daunting challenge, especially in a tight labor market like Singapore with an increasing number of tourist arrivals each year. Moreover, demand for limousine services in the hotel is often volatile; a fair number of guests tend to request limousine services without prior reservation, making both timing and volume of the demands unpredictable for operational planning. Since high quality and yet affordable service cannot be fully provided using a fixed internal fleet of vehicles and drivers, the hotel has to engage external vendors to provide a ``flexible'' buffer of resources available to meet demands in the peak periods. The hotel studied in this paper has a composite fleet structure consisting of over 70 in-house drivers and 34 limousines as its in-house fleet, and 10 other drivers and cars from a contractor as its base fleet. The contractor can also provide additional drivers working in flexible shifts, serving as the hotel's disposal fleet that can be decommissioned when no longer needed, with at least half an hour advance notice.
	
	Managing the delicate balance between demand and supply directly impacts the bottom line of the limousine department. However, because outsourced vehicles need to be activated in time  to avoid ``no vehicle available'' situations, and because demand is erratic, the hotel often erred on the safe side and had an excessive number of vehicles on standby, leading to low productivity and wasted resources. The solution was not sustainable, and hence a more scientific forecasting method beyond intuitive prediction based on experience was needed to forecast demand and manage supply. 
	
	The old  system also relied too heavily on manual tracking of limousine bookings and used different billing methods for the entire fleet of vehicles. Even finding a consistent way to do the reporting proved too difficult, despite numerous attempts and efforts to standardize the format. At the same time, 
	vendor invoicing arrangements involved separate and independent billings to two departments (Transportation and International Marketing). Due to poor communication and lapses in internal controls, the hotel had substantial billing errors. On numerous occasions,  the vendor double-charged the hotel or incorrectly charged one department's job to the other. In addition, many discrepancies in the invoices were unresolved and were pending resolution due to data issues in the limousine booking system. These discrepancies contributed to delayed payments to the vendors.  It was clear that an integrated and automated system was needed to remove the process inefficiencies, and to allow management to  monitor and deploy resources.

\begin{figure}[htbp]
		\begin{center} 
			\includegraphics[width=15cm, height=8cm]{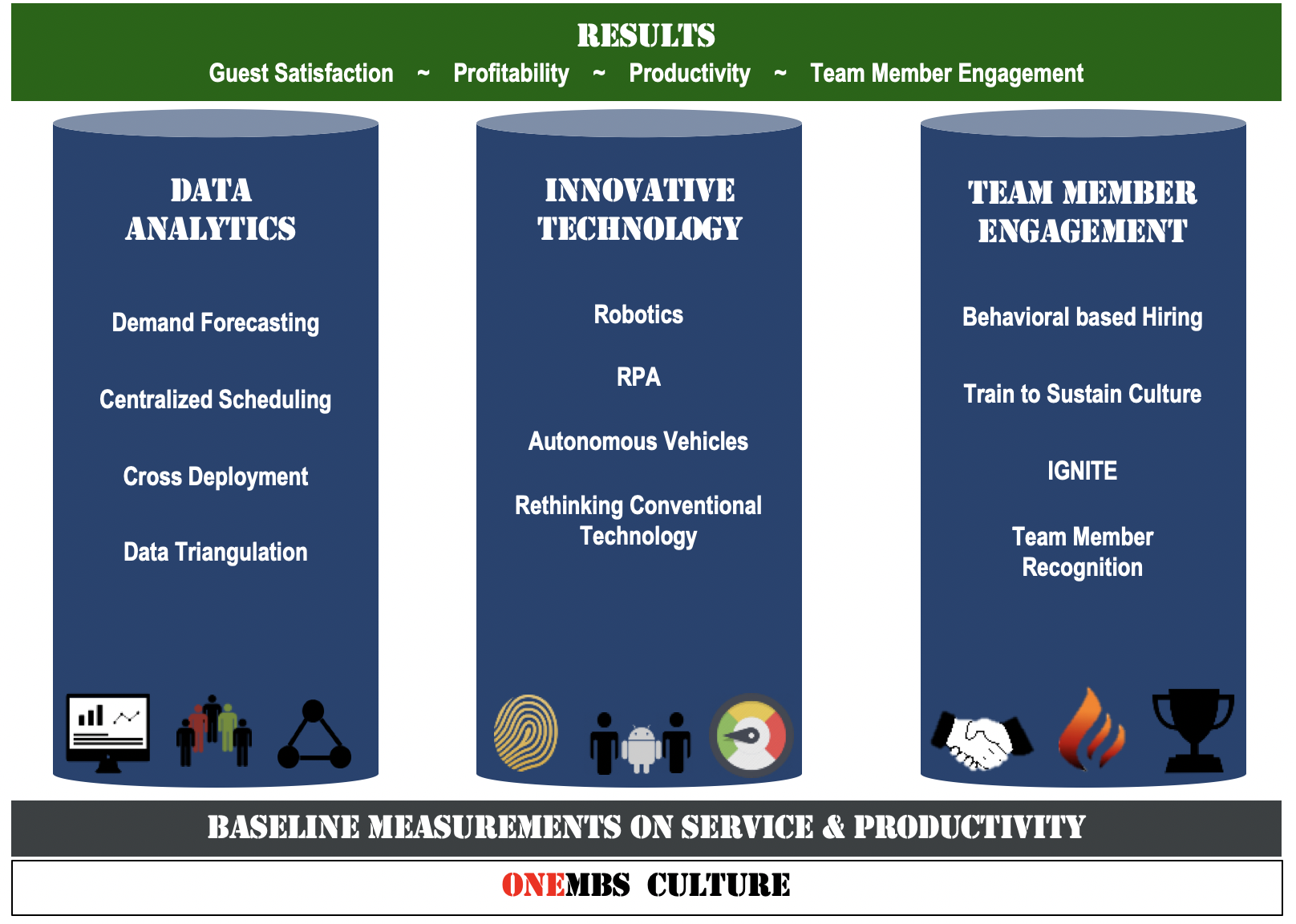}\hspace{0.1cm}
			\caption{Productivity and service strategic roadmap of the hotel in response to Singapore's Smart
				Nation initiative, highlighting the three pillars - data analytics, innovative technology and team member engagement. 
			}
			\label{fig:roadmap} 
		\end{center} 
	\end{figure}

	The study undertaken in this paper is part of the hotel's responses to the Singapore’s Smart Nation initiative, designed to prepare the country for the new digital economy. The hotel has chosen to focus on the strategic roadmap shown in Figure \ref{fig:roadmap},  around the culture of innovation, emphasizing (i) data analytics, (ii) innovative technology, and (iii) team member engagement. Data analytics is at the core of data-based scientific thinking and decision making for the hotel, as represented by its centralized forecasting and scheduling function, which supports property-wide manpower arrangement and cross deployment. This is further enhanced by adopting innovative technologies at both software (e.g. RPA solutions and systems that support real-time monitoring) and hardware (e.g. automatic laundry system and autonomous vehicles) level. The improved productivity enables its staff to better focus on their guests, with many initiatives designed to improve guest satisfaction, which in turn increases its profitability. The hotel also invests significantly in its cultural development by conducting customized training programs and regular engagement sessions with the management team, all of which purposed at shaping a caring and sustainable culture.
	
	Hotel management believes in the value of ``Data over Anecdotes'' and has in recent years established new baseline measurements to focus on reporting cause and effect, in as near to real time as possible. To support these efforts, the hotel has explored and installed people counting and crowd monitoring cameras in numerous locations to generate queue and traffic flow information in real time, has adopted beacon and radio-frequency identification (RFID) technology to track equipment and team members, has added point-of-contact iPad surveys at front desk and parking area, and has developed real-time streaming dashboards using big data services and the Google Cloud Platform (GCP). 
	
	The value of the data captured is further enhanced by robotic process automation, to 
	mimic or replicate a human by carrying out a series of steps that the same person would do, to aggregate or “stitch” together data from multiple sources and systems automatically, and to move data between systems. This effort has culminated in the adoption of software “robots” with configurations and algorithms to automate manual, repeatable tasks.
	
	
	The limousine service provisioning problem is one of the first areas where scientific forecasting principles and data analytics were adopted to solve operational problems in the hotel. Using time series modelling and optimization techniques to predict transaction volumes and manage the demand for and supply of limousines, the output is then refined and polished using expert opinion and human edits. This combined approach allows the hotel to schedule its supply of limousines more effectively to maintain service levels and to increase productivity. 
	
\begin{figure}[htbp]
		\begin{center} 
			\includegraphics[width=16cm, height=8.5cm]{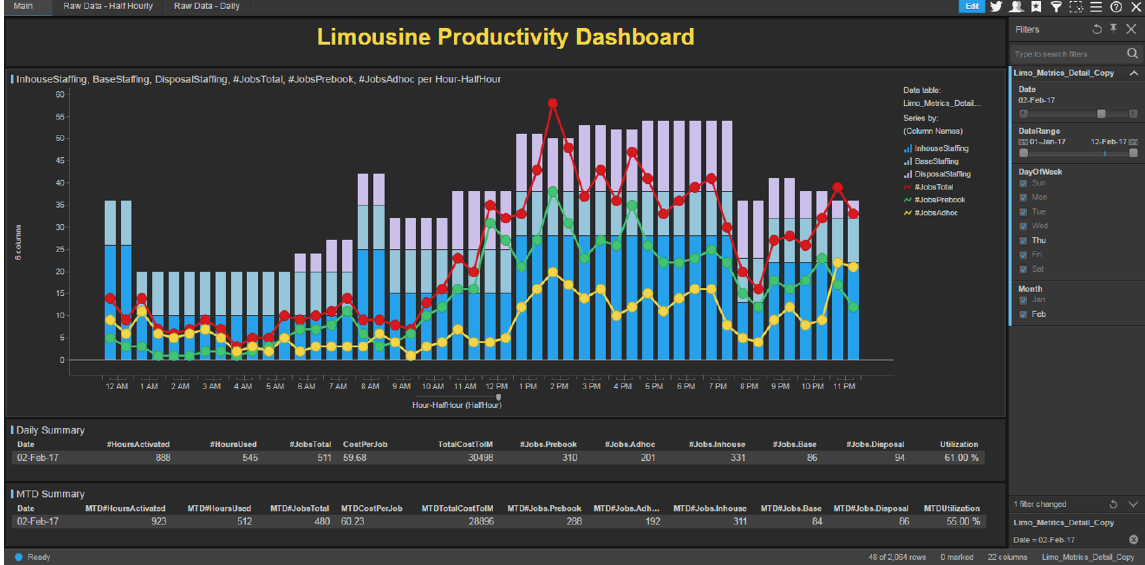}\hspace{0.1cm}
			\caption{Daily managerial dashboard for limousine operations. The hourly demand (red line) consists of pre-booked (green line) and adhoc (yellow line) jobs, while the hourly supply is indicated by the stacked bars from in-house fleet (dark blue bar), base fleet (light blue bar) and disposal fleet (light purple bar). Note that the data are presented at an hourly interval after inflating the job duration for ease of reporting, although an up-to-the-minute picture would be a more accurate depiction of the reality.}
			\label{fig:8} 
		\end{center} 
	\end{figure}

	The project has also revealed cross-deployment opportunities between different operations within the hotel. The project team has developed comprehensive dashboards using Qlikview, Spotfire and R Shiny, which organize all relevant data in a structured manner to visualize and present the information. A daily summary report is also generated for the management to monitor  key metrics such as daily and month-to-date productivity, service level and cost. The dashboard thus aided overall management of the operations and facilitated timely action for improvement. Figure \ref{fig:8} shows the dashboard used on a typical day, displaying the number of ad-hoc and pre-booked jobs, the total number of jobs (in line plots), and the number of in-house, base and disposal staff (in bar plots). 
	
	The underlying engine and decision support system have helped the hotel to make timely and accurate decisions, with seven figure savings while achieving 99.97\% vehicle availability for 105,000 limo transfers in the year of 2016. The following sections first highlight the encountered issues such as robust forecasting and business process automation, followed by an in-depth review of the adopted solution and its implementation across the hotel.

	\subsection{The Issues}
	
	The limousine service provision problem mainly revolves around managing the delicate balance between supply of vehicles and demand from guests, providing just enough vehicles at the right time. To achieve this, there are three challenges:
	\begin{itemize}
		\item Accurate forecasting of demands. The operations team used to assess the weekly demand qualitatively, based on recent demand patterns using three-period moving average and future pre-booked jobs. Forecasting the demand for limo drivers and vehicles on a particular day is difficult because the usage of the vehicle can vary widely depending on the nature of the trips. Table \ref{tbl:events} shows sample events related to the trips and the various attributes that are used in developing the forecast. In an integrated resort that offers MICE (meetings, incentives, conferences and exhibitions) solutions, the demand for limousine services is often volatile. The category of an event (wedding, business meeting, etc.) and the number of attendees for the event affect the duration of the trip. The model needs to anticipate the number of such requests, and also the duration of usage, to arrive at a cumulative hourly demand for limousines in the department. The model must also utilize historical data (from past usage) and incorporate the implicit expert knowledge on the impact of future events to build the final demand forecast.
	\end{itemize}

	\begin{table}[htbp]
		\begin{center}
			\resizebox{0.95\textwidth}{!}{
				\begin{minipage}
					{0.95\textwidth}
					\caption{Sample events and their attributes}
					\label{tbl:events}
					\begin{tabular}{p{3cm}p{2cm}p{2cm}p{2cm}p{3cm}p{2cm}}
						\hline\hline
						Event & Start Time & End Time & Status & Category & Attendance 
						\\  \hline
						Weekly Planning Meeting & 02/12/2016 10:00 am & 02/12/2016 02:00 pm & Definite & Internal Event & 40
						\\ \hline
						Wedding Dinner & 30/12/16 08:00 am & 30/12/16 11:00 pm & Definite & Wedding & 210
						\\ \hline
						Church Service & 24/12/16 08:00 am & 26/12/16 08:00 am & Tentative & Local Business & 5000 
						\\ \hline
						Exhibition & 20/11/16 08:00 am & 23/11/16 08:00 pm & Definite & Local Business & 2000 
						\\ \hline
						Dinner \& Dance & 25/12/16 08:00 am & 25/12/16 08:00 pm & Tentative & D\&D & 1000 \\
						
						\hline\hline
					\end{tabular}
				\end{minipage}
			}
		\end{center}
	\end{table}
	
	\begin{itemize}
		
		\item Sufficient and proper deployment of limousines subject to operational constraints. The weekly deployment of the limited resources (drivers and limousines) must cover the predicted demand, subject to certain operational constraints, while not incurring additional wasted resources. For example, a driver cannot work for more than 12 hours in a single shift, and the system can only deploy up to certain number of shifts in a single day.  Such operational constraints have led to a total of four shifts for in-house drivers in the hotel, starting at 8am, 11am, 1pm and 9pm respectively and each lasting 12 hours for four days in a row, followed by a four-day rest. Combined with  (say) 10 fixed base fleet drivers from the subcontractor who are available 24/7, the difference between supply and demand can be easily derived.  Figure \ref{fig:buffer} depicts a situation where the total number of cars (in-house plus base fleet from vendor) falls short of the forecast demand forecast at 11am, 12pm, and from 4pm to 8pm. Additional disposal vehicles need to be deployed to guarantee sufficient coverage of the forecasted demand, subject to various operational constraints as follows:
		
		\begin{itemize}
			\item A minimum of three hours, and a maximum of eight hours per shift, is required for each activation of disposal vehicle. 
			\item The earliest activation starts at 5am, and the number of activated disposal vehicles should not exceed 60 in each shift.
			\item At most six shifts should be scheduled daily.
		\end{itemize}
		As the standard roster planning system of the hotel cannot handle these constraints all together, the weekly schedule for disposal fleet was often produced by operational managers based on personal preferences and experiences, leading to inconsistent, unstable and sub-optimal solutions. A customized solution based on mathematical programming is thus desired.   
		\begin{figure}[htbp]
			\begin{center} 
				\includegraphics[width=15cm, height=8cm]{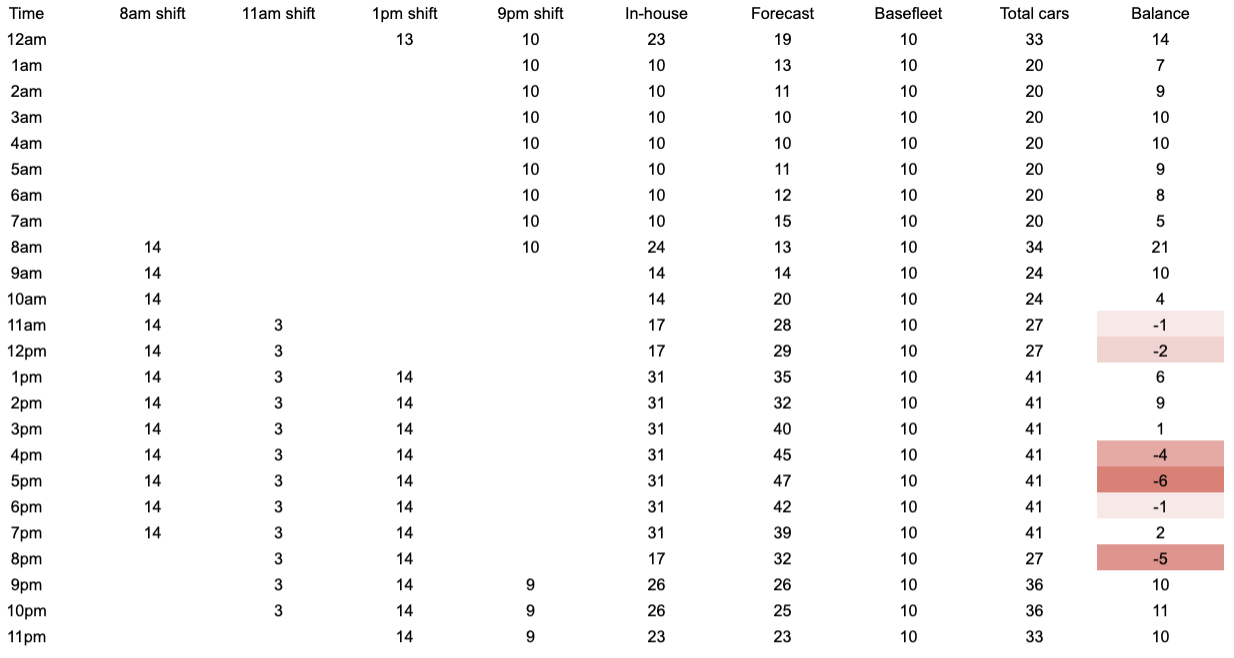}\hspace{0.1cm}
				\caption{Example of deriving hourly buffer using in-house and base fleet against forecasted demand. Hours with negative balance need to be served with the disposable fleet of drivers, subject to operational constraints}.
				\label{fig:buffer} 
			\end{center} 
		\end{figure}

		\item Continuous reporting and monitoring of operations. Limousine operations require near real time monitoring of available supply and demand. To better anticipate upcoming business volumes for the next few hours, the operations team needs constantly to generate status reports indicating the current operational status including number of pre-booked jobs, vehicles activated and available buffer, and to determine if additional vehicles need to be activated. This 3-hourly process, however, takes around half an hour to run, as it requires manually retrieving and processing data from different systems  including the business management and roster systems. Further data processing in a spreadsheet is also required to produce the hourly metrics. Often the report generating process is interrupted by unexpected operational challenges which require immediate attention. In addition, the report doesn't consider historical distributions of demand, thus giving no indication as to whether some of the disposal vehicles could be de-activated to save cost. 
		
	\end{itemize}
	
	In this paper, an integrated approach is introduced to solve the limousine scheduling problem for the hotel. In order  to accurately forecast the demands so that a high service level can be achieved, the objective is to focus on minimizing the stockout situations (i.e. no-vehicle-available cases), while keeping cost low to avoid unnecessary outlays. To do this, a combination of statistical tools, spreadsheet optimization and data analytics is used to support operations in the following three ways. 
	\begin{itemize}
		\item A demand forecasting model was implemented effectively to predict business volume, together with an optimization model to schedule the corresponding shifts of drivers and vehicles. 
		\item Next, a mission critical bot was developed to update operational status on an hourly basis and to provide recommendations on the time and number of disposal vehicles to be stood down. 
		\item Finally, a centralized managerial dashboard was developed to display key daily performance indicators, thus enabling effective performance monitoring, scenario-based planning and decision-making. 
	\end{itemize}
	
	Note that several papers have approached similar manpower staffing problem from different perspectives. For example, \cite{limo2010} introduced a Fleet Management System (FMS) which supports efficient scheduling and management of vehicles by utilizing open-source technology and robust scheduling procedures, while \cite{taxi2020} illustrated how taxi services can be optimized using key performance metrics such as number of pick-ups, customer waiting time, and vacant traveled distance. However, none has covered the use of statistical and optimization tools and applications in analyzing high-frequency demand data with complex seasonal patterns and multiple concurrent events, the case often encountered in the hospitality industry.

	\section{Solution Overview and  Implementation}\label{sec:DataAndMethodology}

	\subsection{Demand Forecast with Time Series Model}\label{sec:Data} 
	
	A valid limousine job consists of multiple attributes, including start time, end time, and job type. The job type could be airport arrival or departure, single trip, round trip, or disposal at guest's will, each taking a different amount of time to complete. Different types of jobs and their typical job duration, assumed to follow an empirical normal distribution, are shown in Table \ref{tbl:duration_assumption}, along with the standard deviation often observed by operations team. Note that this table is used to establish the duration of jobs with missing or incorrect information in the business management system, although a further refinement based on hour of the day would better capture the hourly sensitivity across the day.
	
	\begin{table}[htbp]
		\centering
		\begin{center}
			\caption{Typical duration for each type of job in minutes}\label{tbl:duration_assumption}
			\begin{tabular}{ccc}
				\hline\hline
				Job Category  & Mean & Standard Deviation \\
				\hline
				Round Trip & 95 & 20 \\
				Single Trip & 70 & 15   \\
				Airport Arrival  & 90 & 20 \\
				Airport Departure & 50 & 10 \\
				Ferry Arrival & 90 & 20 \\
				Ferry Departure & 40 & 10 \\ 
				Malaysia Transfer & 360 & 60 \\
				\hline\hline
			\end{tabular}
		\end{center}
	\end{table}

\begin{figure}[htbp]
		\begin{center} 
			\includegraphics[width=12cm, height=7cm]{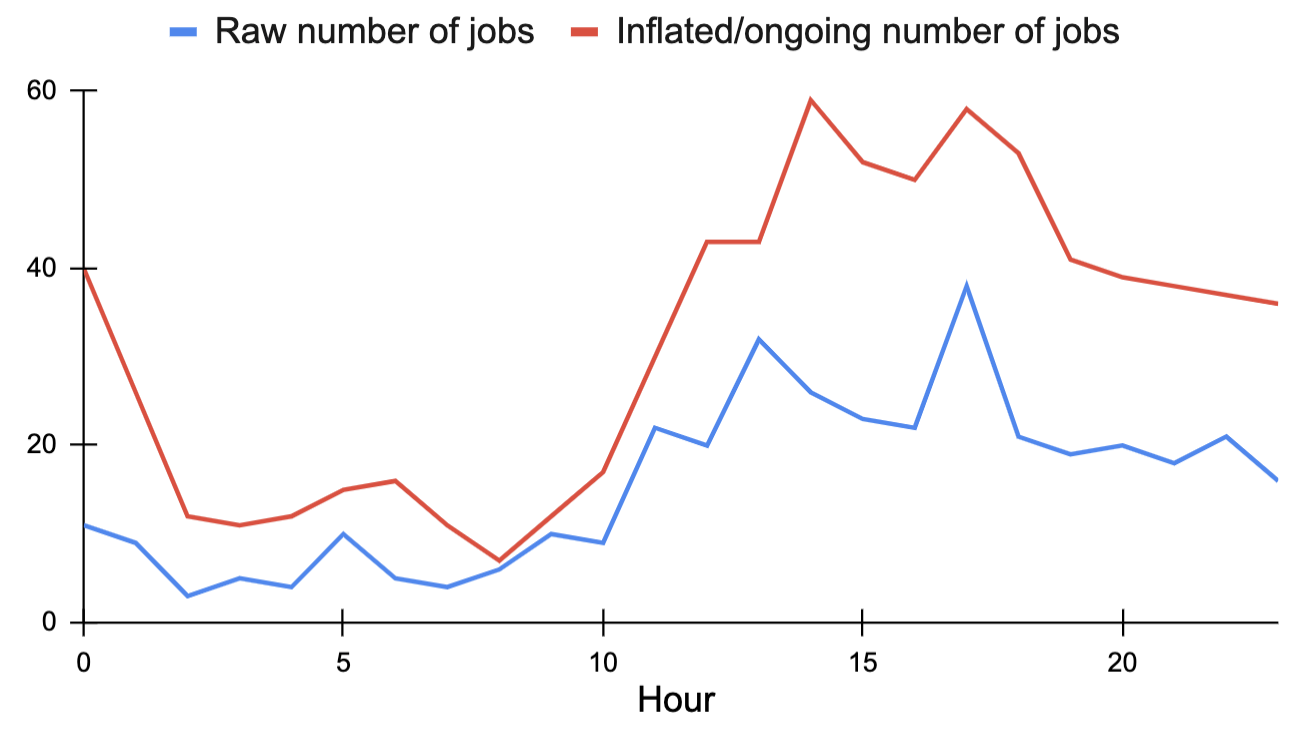}\hspace{0.1cm} 
			\caption{Deriving the true business volume via inflating the job duration according to different type of job. The inflated number of job (red line), an indication of ongoing business volume, is generated by taking into account the duration of the raw jobs (blue line)}.
			\label{fig:1} 
		\end{center} 
	\end{figure}

	In order to estimate the number of ongoing jobs per hour, or equivalently, the number of cars required per hour, it is necessary to account for the job duration, since each job may last more than an hour. Figure \ref{fig:1}  illustrates the original volume and the ``inflated" ongoing jobs on an hourly basis, which represents the true business demand for limousines for each hour.
	
\begin{figure}[bht]
		\begin{center}  
			\includegraphics[width=12cm, height=7cm]{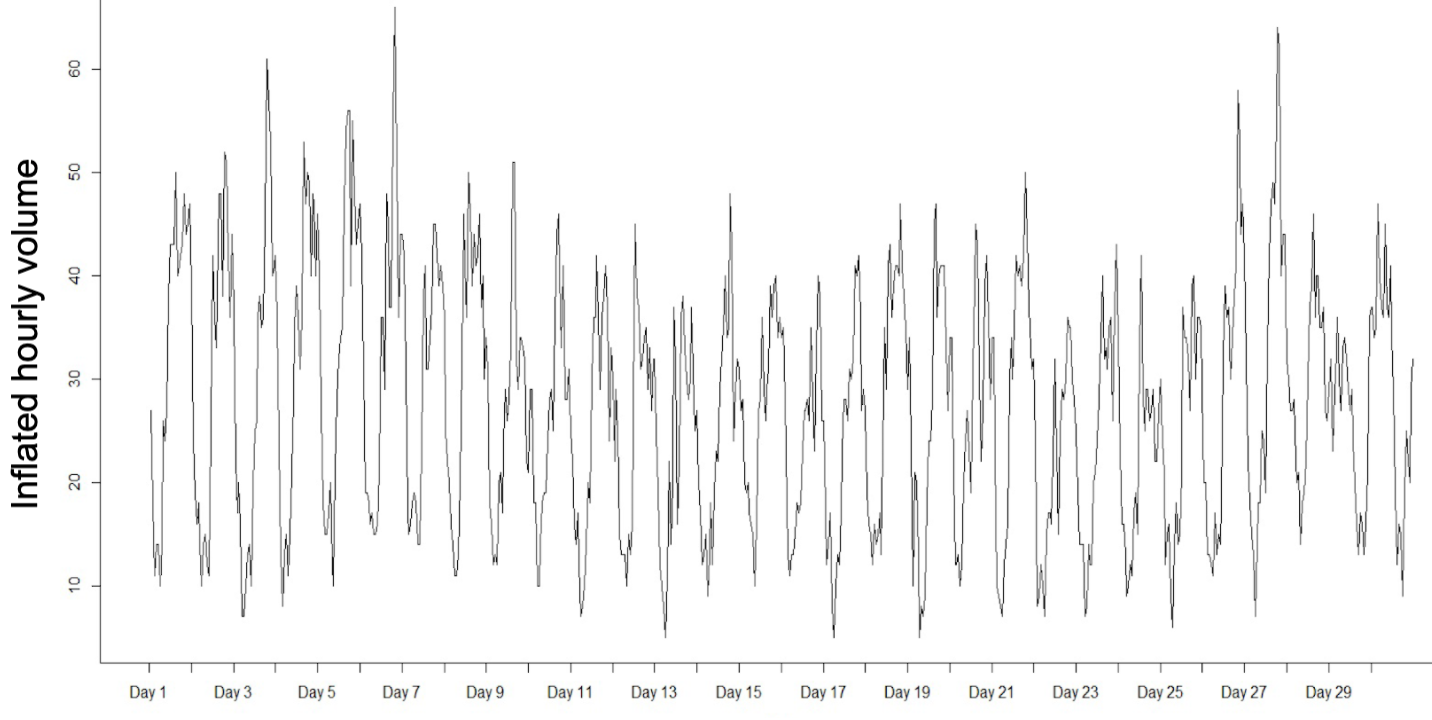}\hspace{0.1cm} 
			\caption{Inflated hourly volume over 30 days.}
			\label{fig:2} 
		\end{center} 
	\end{figure}

	Figure \ref{fig:2} shows the hourly number of ongoing jobs (after inflation) over a 30-day horizon.  For limousine services which operate around the clock, the business demands can exhibit multiple seasonality patterns. For example, the hourly number of guests requesting limousine services has a short term daily seasonal pattern, a weekly seasonal frequency, and a longer annual seasonality pattern. To forecast demand in this problem, the underlying trend and seasonality  need to be uncovered from the data.
	
\begin{figure}[htbp]
		\begin{center} 
			\includegraphics[width=12cm, height=7cm]{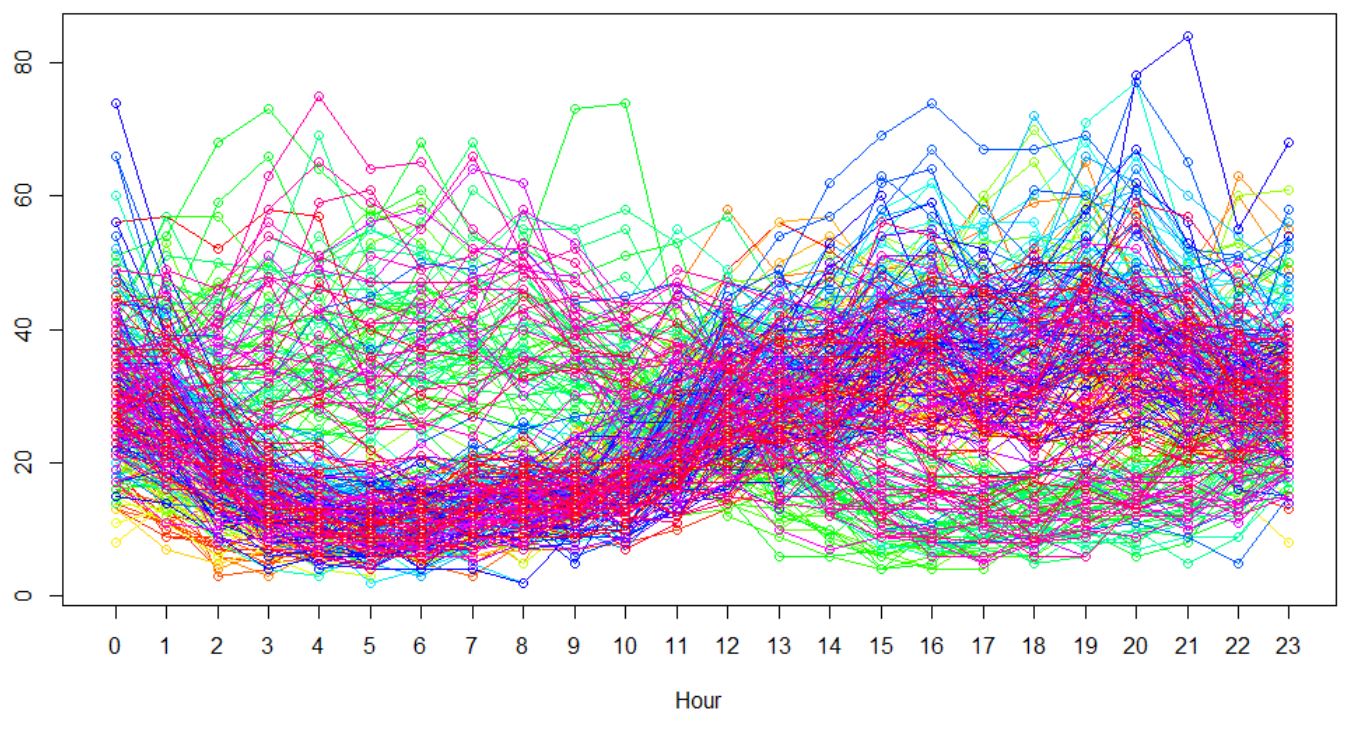}\hspace{0.1cm} 
			\includegraphics[width=12cm, height=7cm]{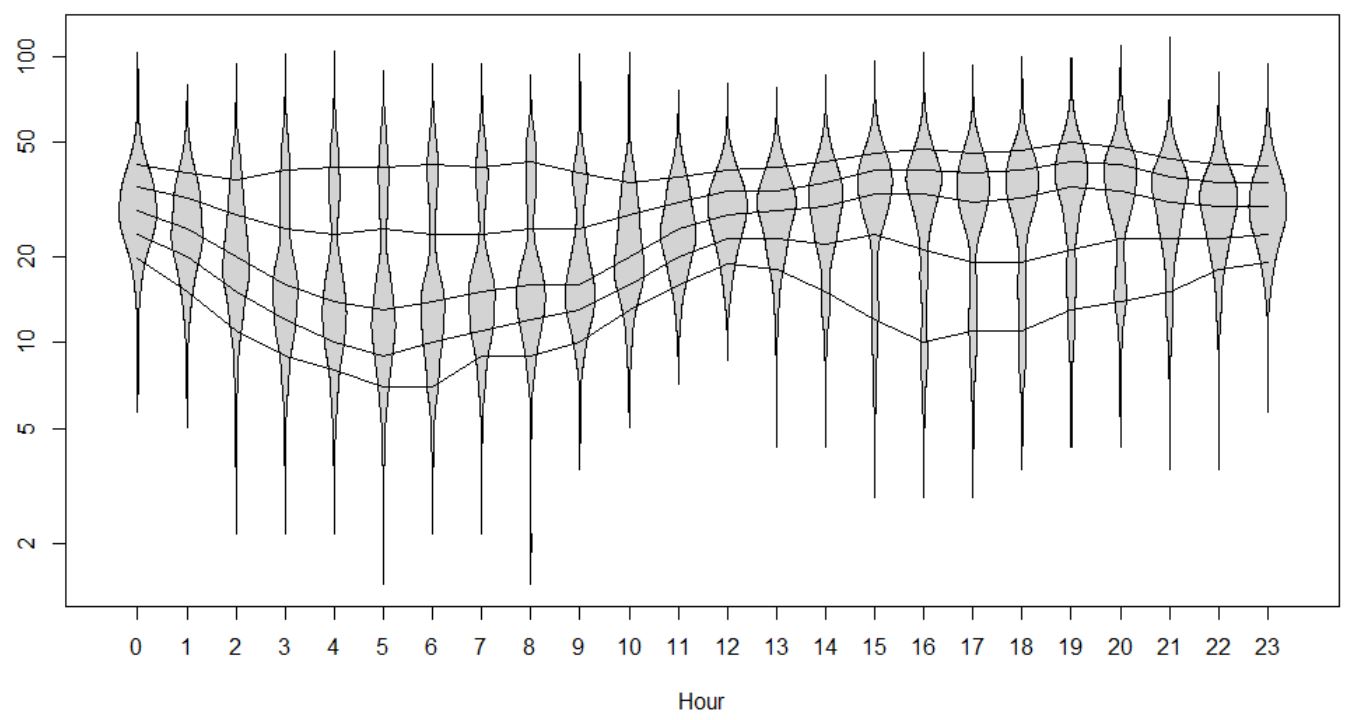}\hspace{0.1cm} 
			\includegraphics[width=12cm, height=7cm]{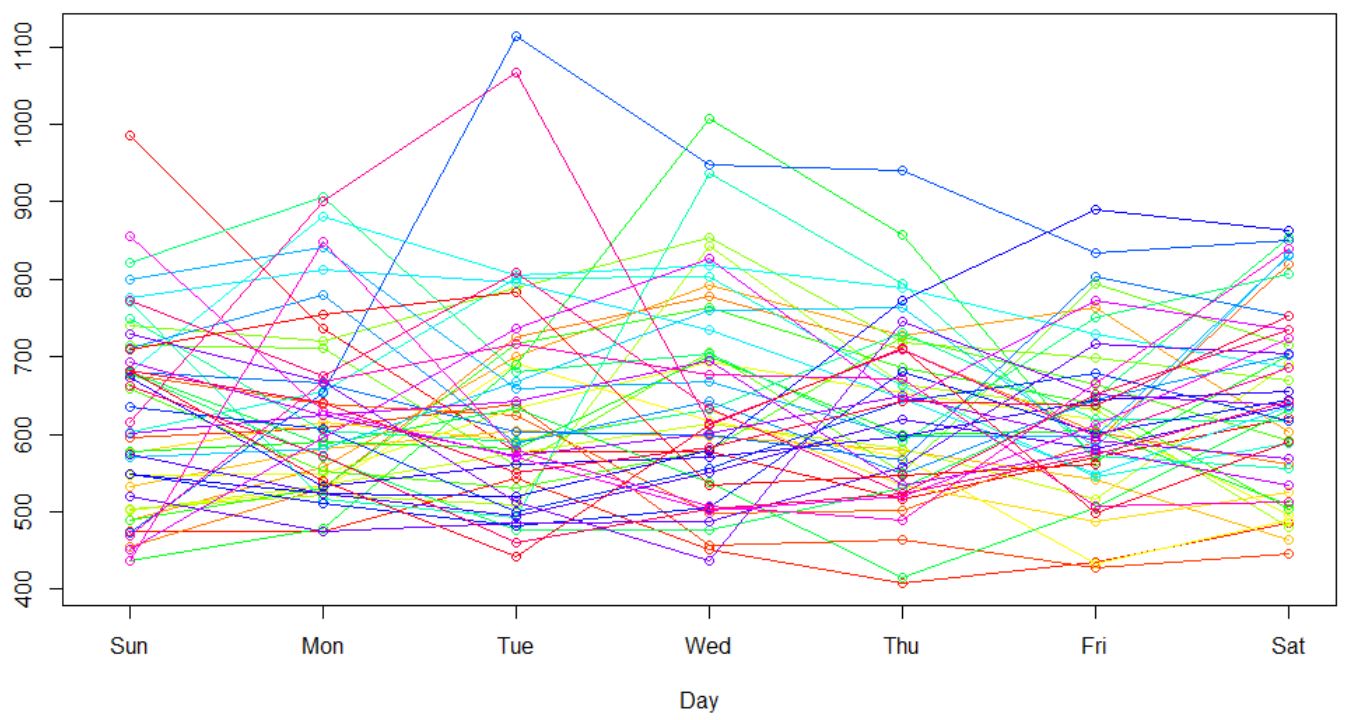}\hspace{0.1cm} 
			\caption{Inflated hourly volume across 2017, where each line in the first graph represents one day's inflated volume. In the second graph, each line represents 10\%, 25\%, 50\%, 75\% and 90\% quantile respectively.  Peak volume above the 90\% percentile is usually due to events such as Jay Chou concert. In the third graph,  each line represents total inflated volume for each day of week}
			\label{fig:seasonal_1} 
		\end{center} 
	\end{figure}

	To gain a better understanding of the seasonal component in the time series, the inflated hourly volume across the whole year of 2017 is visualized in both line chart and bean plot (see Figure \ref{fig:seasonal_1}). Demand volume typically starts to increase gradually from around 12pm, peaking between 5pm and 10pm, with considerable volume arising towards the end of the day. These peaks are consistent with our observation that most guests tend to engage in entertainment activities and explore the night life at the hotel and in Singapore, leading to higher volume in the evening than in the morning hours. The hour-of-the-day seasonality is fairly consistent across the whole period, except for certain periods when external factors such as major events in the city (e.g. a concert by a well known singer)  result in atypical volume on a certain day. Moreover, by aggregating daily volume as a whole and plotting aggregated volume by each day of week, a certain level of seasonality is also observed as shown in the last graph in Figure \ref{fig:seasonal_1}. In general, business volume from Friday to Sunday is heavier than from Monday to Thursday, as more guests  engage in leisure activities over the weekend.

	The common seasonal models used in practice include the well-known Holt-Winters’ additive and multiplicative methods, and various extensions to include more seasonal components. The Holt-Winters’ method is extended by \cite{forecast2010} to handle non-integer seasonality and calendar effects, or time series with non-nested seasonal patterns. This approach includes a Box-Cox transformation, ARMA errors and multiple seasonal patterns as follows. The model is called TBATS, with T standing for trigonometric, B for Box-Cox transform, A for ARMA errors, T for trend, and S for seasonal components. In this model, a trigonometric representation of seasonal components based on Fourier series is introduced to reduce the number of parameters that require estimation. Details on the model are included in the appendix, and the derivation is available in the references therein. Implementation of this model is based on the ``forecast'' package using R, after smoothing out outlier observations usually caused by events, public holidays or system recording error. The outliers are identified as jobs with duration above the 95\% percentile of full empirical distribution on job duration. Since removing the outlier observation would disrupt the seasonality in time series, a typical value from a normal day, for the same hour of the day and day of the week, is selected from the historical data to replace and therefore smooth out the outliers.

	This approach performs well for our problem. As an illustration, the data between January and September in 2016 are used as a training set to assess the performance of this approach in Oct 2016. The average results are tabulated in Table \ref{Table: 1}. By comparing the two methods with respect to three different metrics - root mean squared error (RMSE), mean absolute error (MAE) and mean absolute percentage error (MAPE), it is observed that the forecasting model proposed appears to dominate the moving average method in all dimensions, measured at a daily level for one-day-ahead forecasts across all the days in test set.
	
	Compared to the previous forecasting model using a simple three-period moving average, the superiority of the enhanced model is demonstrated in the following three aspects. First, better test set performance in all three metrics. This is due to a further exploitation of structural components like trend and multiple seasonalities in the time series over a longer horizon. Second, automated weekly forecasts. The previous forecasting exercise requires manual processing over spreadsheet every week, suffering from operation risks such as incorrect entry. The automated forecasting system thus offers a more reliable and consistent output for consumption. Third, as mentioned in the later part of the paper, the system gives a visual representation of both forecasted and historical value by hour of the day and day of the week, supporting critical decisions such as determining if the current stock of vehicles needs to be boosted or reduced. 
	
	Since new data are constantly generated, a monthly re-training task is scheduled to re-estimate the model coefficients and re-align to the latest changes in data. However, the refreshed TBATS model occasionally suffers from overfitting and generates unstable and excessive hourly forecasts. To avoid such circumstance and further stabilize the automated forecasting system, a separate seasonal moving average model was added in an equally weighted ensemble fashion from 2017 onwards. This results in more operationally consistent and reliable forecasts, yet the performance is on par with the standalone TBATS model on normal days.

	\begin{table}[htbp] 
		\centering 
		\caption{Comparison of average one-day-ahead daily out-of-sample performance over one month.} 
		\label{Table: 1} 
		\begin{tabular}{p{6cm}p{2cm}p{2cm}p{2cm}} 
			\hline \hline 
			Forecasting Model & RMSE & MAE & MAPE 
			\\ \hline 
			Previous model based on three-period moving averages & 9.311 & 7.1 & 0.306 
			\\ \hline 
			Multi-seasonal time series model & 8.297 & 6.206 & 0.249
			\\ \hline \hline 
		\end{tabular} 
	\end{table}

	\subsection{Collaborative Event-based Forecast}\label{sec:FeatureConstruction} 
	Although mathematical forecasting approaches can lead to reliable demand forecasts by 
	extrapolating regular patterns in time series, unpredictable events that do not appear in historical data can reduce the usefulness of mathematical forecasts for demand planning. Since forecasters have partial knowledge of the context and of future events, it is important to group and structure the fragmented expert knowledge to be integrated into final demand forecasts. In this regard, the approach proposed by \cite{combine2011} is followed. The baseline mathematical forecast is adjusted by structured and combined knowledge from different forecasters, thus leading to a more accurate and practical forecast. 
	
	\begin{figure}[htbp]
		\begin{center} 
			\includegraphics[width=15cm, height=10cm]{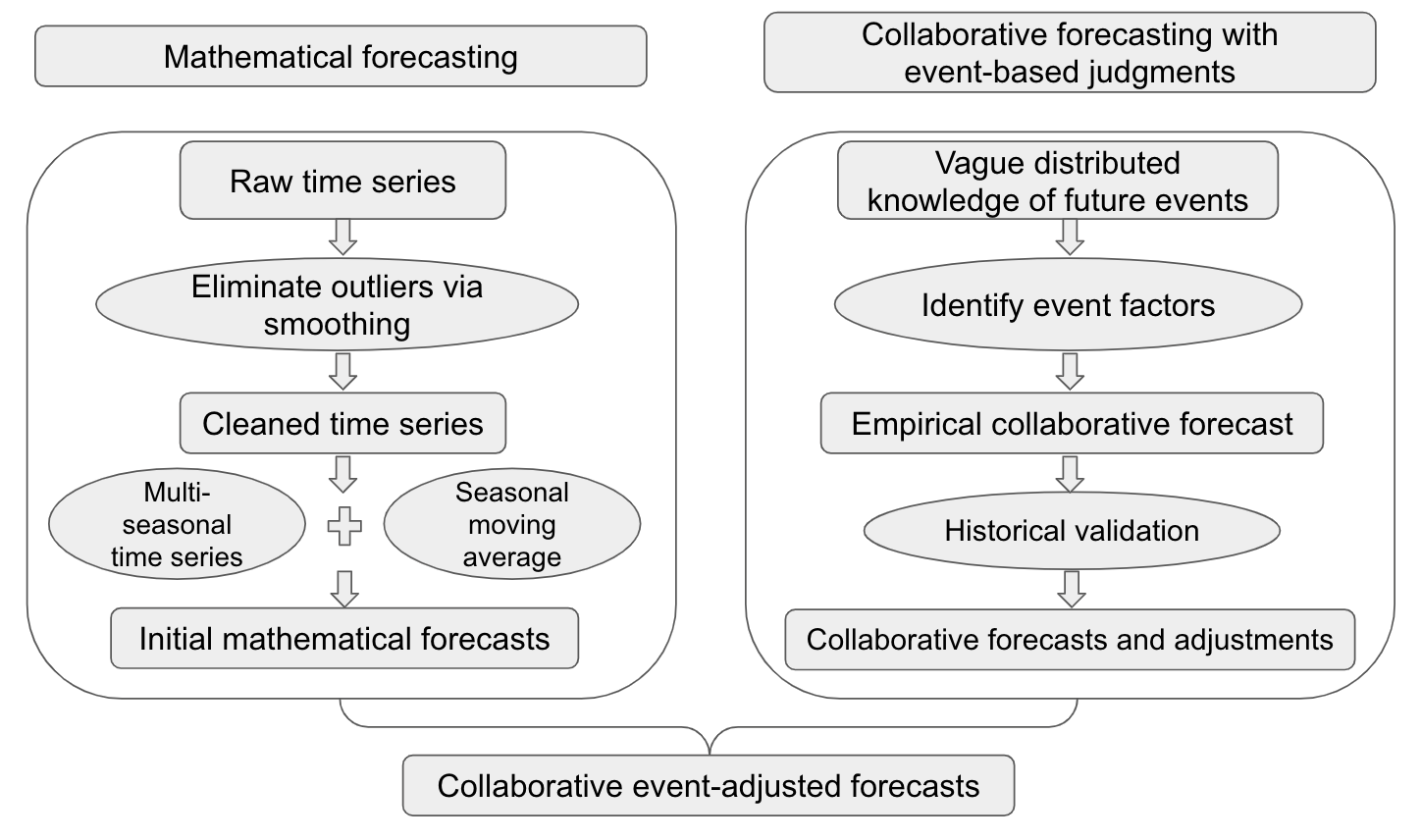}\hspace{0.1cm} 
			\caption{Flow chart of collaborative event-adjusted time series forecast. The initial baseline forecast is generated using an ensemble of multi-seasonal time series model and seasonal moving average model based on smoothed historical data. The output from mathematical forecasting is then editted, often increased, by operational experts based on relevant event factors, producing the final collaborative event-adjusted forecasts.}
			\label{fig:5} 
		\end{center} 
	\end{figure}

	To better understand the impact of different events on the baseline model, a thorough analysis was jointly conducted by forecaster and planners/controllers on the quantitative impact from events or public holidays, including the details of events such as event type, number of attendees, duration, and internal operational status on a particular day. Recall that Table \ref{tbl:events} shows some of the events and associated attributes, which may also differ from their actual value. For example, a wedding dinner may end later than originally scheduled, while multiple weddings on the same day may result in a bigger shift in demand pattern as it may be a public holiday or an auspicious day in the calendar. In addition, for big events in the city with over 5000 anticipated attendees, the actual attendance is likely to fluctuate considerably. This  changes the demand for limousine services at the hotel. Since it is very difficult for the forecasting model to capture the dynamics in a demand pattern driven by different events and other factors, it is observed that the multi-seasonal time-series model is more likely to be inaccurate during days with big events. Therefore, being aware of this reference information on future events and corresponding uncertainties, operational staff will evaluate the impact of such events by observing the demand pattern during similar days in the past, and quantify the additional amount to be added on top of the aforementioned forecasting output as a measure of empirical adjustment. 
	
	The collaborative event-adjusted forecasting process is illustrated in Figure \ref{fig:5}. First, the raw time series data is carefully cleaned by smoothing out the potential outliers using historical average, presenting an event-free picture of the demand pattern. Second, an ensemble of multi-seasonal time series model and seasonal moving average model is applied to generate the initial mathematical forecasts, followed by operational perusal and potential edits from experts based on relevant internal and external events or public holidays. Once finalized, the collaborative event-adjusted forecasts are then submitted for weekly planning of fleet schedule.

	\begin{figure}[htbp]
		\begin{center} 
			\includegraphics[width=13cm, height=8.5cm]{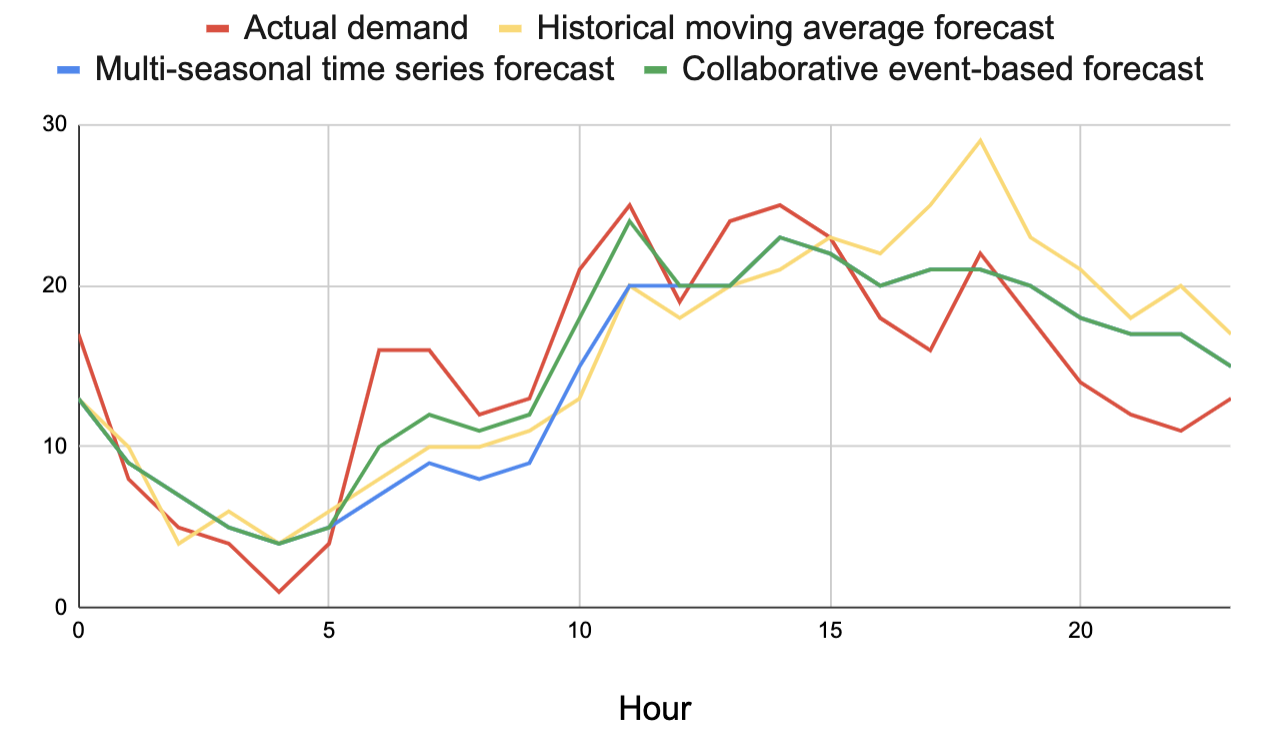}\hspace{0.1cm}
			\caption{Comparison of different forecasts on October 5 2016. The multi-seasonal time series forecast (blue line) is able to better track the actual demand (red line), especially during the later half of the day (overlapped with green line), compared with the historical moving average forecast (yellow line). The peak in the actual demand from 5am to 12am, driven by a big group check-in, is partly captured by the collaborative event-based forecast (green line), which is an improvement compared with the blue line. The additional gap between red and green lines is also used to update the operational knowledge on the impact of similar events, leading to better edits in the future.}
			\label{fig:6} 
		\end{center} 
	\end{figure}

	For instance, the operations team knew that there was a big group checking in at the hotel on the morning of October 5 2016. This knowledge was not accounted for in the mathematical model. Based on past data, the operations team was able to provide a rough estimate of additional jobs needed for such a group event. As illustrated in Figure \ref{fig:6}, the improvement in accuracy using the revised forecast is obtained by adding on-the-ground knowledge. The revised composite forecast fits the actual demand better than the mathematical model. Table \ref{Table: 2} quantifies the improvement in RMSE, MAE and MAPE, using the revised forecast, for the same data used in Table \ref{Table: 1}.  The composite forecast proves to be better in predicting future demand than single model-based forecasts, especially during days with big events in the city, i.e. China's Golden Week holiday in October.

	\begin{table}[htbp] 
		\centering 
		\caption{Comparison of average daily out-of-sample performance for the three models over one month} 
		\label{Table: 2} 
		\begin{tabular}{p{6.5cm}p{2cm}p{2cm}p{2cm}} 
			\hline \hline 
			Forecasting Model & RMSE & MAE & MAPE 
			\\ \hline 
			Previous model based on three-period moving averages & 9.311 & 7.1 & 0.306 
			\\ \hline 
			Multi-seasonal time series model & 8.297 & 6.206 & 0.249
			\\ \hline
			Collaborative event-based model & 7.791 & 5.78 & 0.239 
			\\ \hline \hline 
		\end{tabular} 
	\end{table}

	\subsection{Optimal Resource Scheduling} 
\begin{figure}[h]
		\begin{center} 
			\includegraphics[width=16cm, height=9cm]{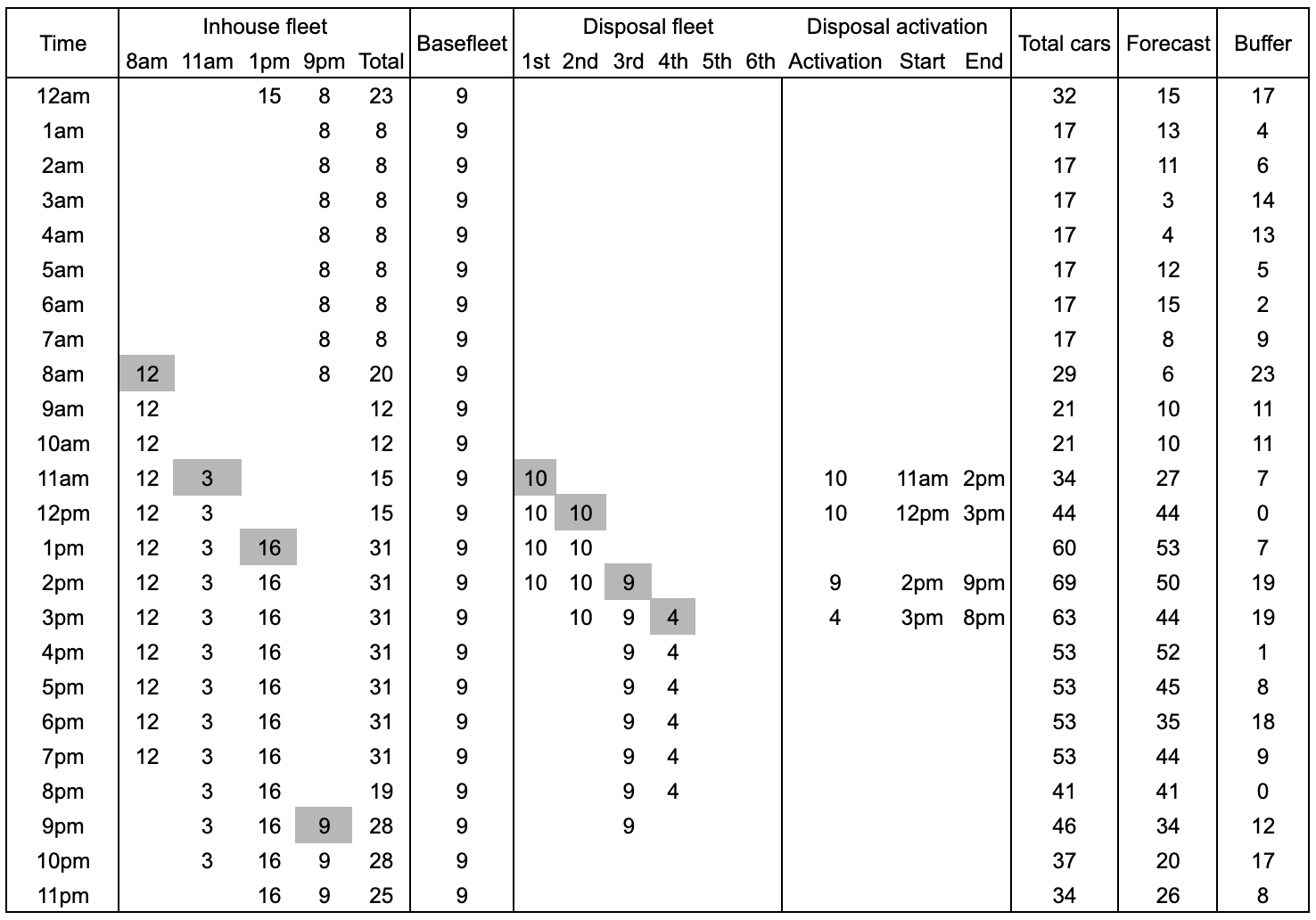}\hspace{0.1cm}
			\caption{An example of the problem solved. 5am is used as the start of the planning horizon, and wrap around midnight into the next day. The shaded cells for shifts correspond to the starting time of the shifts. Note how the balance across different hours of the day is maintained above or equal to zero by the additional support from disposal activation. The disposal activation details, namely activation number, start and end time, are then submitted to the vendor one week in advance.}
			\label{fig:limoexample} 
		\end{center} 
	\end{figure}

	In view of the surge in demand due to events and ad-hoc, last-minute requests, the in-house and base fleets alone are insufficient to cover the demand at all hours of the day, thus necessitating the need for additional support from the disposal fleet. As an integral portion of supply to ensure a high service level, the disposal fleet is also the most expensive fleet due to its flexible schedule. Prior to the adoption of a data driven approach to forecasting and shift optimization, disposal activation was based on gut feel, and often resulted in either under-utilization or a shortage of vehicles. Figure \ref{fig:limoexample} shows a sample problem; the forecast total car column is derived from the  proposed forecasting model. Note that the historical peaks in hourly demand, observed in days with similar characteristics, are also considered, so as to guide the managers to  stand down unnecessary ad-hoc hires, in case the demand is lower than the expected volume. 
	
	The next section shows how the number of disposal vehicles to be activated could be determined automatically using a mathematical programming approach. The section for disposal activation in Figure \ref{fig:limoexample} used to be manually filled in by operational managers based on experience and gut feel. Automating this task is challenging, in part because capturing the constraints in the problem using the natural decision variables (how many cars to reserve at each hour in each shift) is difficult, as the consecutivity and uniformity constraints need to be modelled (i.e., each shift has the same number of drivers for each hour in the shift). Instead, the scheduling constraints are incorporated using a set covering formulation. 
	
	Let $l(j)$ denote the length of shift $j$, and let $s(j)$ and $e(j)$ denote the start and end time of shift $j$.  The set of feasible shift patterns, denoted by ${\cal S}$, is generated in advance. Let $x_{j}$ denote the number of drivers activated for shift $j$. Let $y_j=1$ if shift $j$ is activated and 0 otherwise. Using a mathematical programming approach, the problem is re-formulated with an objective to minimize the total activation hours, i.e. $\sum_{j\in {\cal S}}l(j)x_{j}$, such that the total activated cars are sufficient to cover the forecasted demand. In addition, there are maximum six times of activation for disposal vehicles, and each activation can only accommodate up to 60 vehicles. These constraints are per the service level agreement between the hotel and its vendor. More details on the formulation are available in the appendix. Note that there are at most 19 slots to activate the drivers (from 5am to 11pm), and 6 possible shift lengths (from 3 to 8 hours), and hence at most 114 possible shift patterns. The solution of this formulation can be obtained using a spreadsheet optimization solver.
	
\begin{figure}[htbp]
		\begin{center} 
			\includegraphics[width=16.5cm, height=9cm]{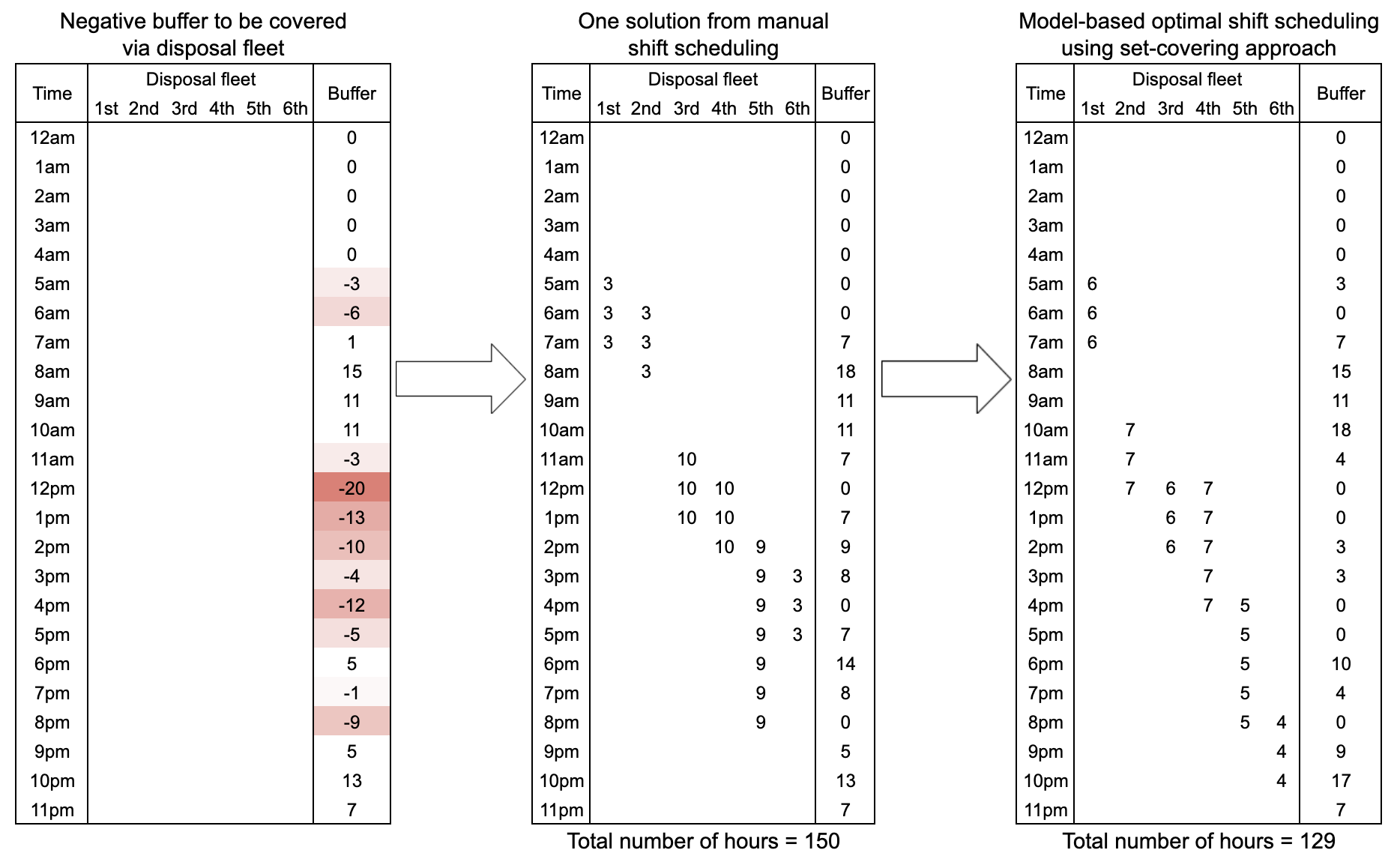}\hspace{0.1cm}
			\caption{Comparison of manual solution versus model-based optimal solution. The optimal solution uses 21 hours less of driver time and yet ensures all buffers are covered, thus being a more preferred schedule.}
			\label{fig:7} 
		\end{center} 
	\end{figure}

	Figure \ref{fig:7} illustrates the scheduling process of a normal day and compares the manual solution with model-based optimal solution. The result shows that the optimal solution uses 21 hours less of driver time and yet ensures all buffers are covered, thus being a more preferred schedule. The model-based scheduling is performed for each day of the week one week in advance, and the solution time is less than 2 seconds for each day using OpenSolver in Excel.
	
	\subsection{Efficient Resource Planning via Process Automation}\label{sec:PerformanceMeasure} 
	
	To strike a proper balance between supply and demand, it is important to equip operational staff with the right data at the right time. Recall that the status report, which givess the supply and demand for the next 12 hours, used to take half an hour if generated manually. Automating this process requires finding solutions to  two challenges - data availability and processing complexity. It is a relatively trivial task to replicate the manual data stitching process using a scripting language like R, so that the latter two components of the usual ETL (Extract, Transform, Load) routine are automated. The main challenge, however, lies in extracting data from operational systems.
	
	The demand data are stored in LIS, a web-based application, while the manpower data are kept in Virtual Roster (VR), a standalone application. Both vendor systems require authentication and proper navigation to locate and download the data needed to produce the status report. The navigation and downloading process could take up to 10 minutes depending on network stability and system loading capacity. Logging in again is necessary should the systems time out during a session.
	
	The significant bottleneck for data availability not only prolongs the report generation time, but also prevents the operations team from having an up-to-the-minute understanding of the ongoing operations, since the status report only conveys the picture half an hour ago. To address this challenge, the project team invested a lot of time to understand the features of the two systems and came up with a two-pronged approach, via a bot, to access the relevant data. For LIS, a self-contained script was developed using R to establish a valid session upon authentication, enter certain parameters such as date range, and download the relevant data, all completed in the backend without manual intervention. For VR, a Python script was developed to download data by performing mouse moves and keystroke entries according to a pre-configured procedure. The key to this approach lies in knowing the exact location to click on the screen and the keys to strike in each step, along with the interval to wait between steps. Although these automated processes work properly most of the time, occasional breakdown still occurs due to unforeseen factors such as network failure or unwanted use of dedicated workstation hosting the system. To overcome this, a separate early warning system is also developed to monitor the execution status and alert the relevant person should a failure occur.
	
\begin{figure}[htbp]\label{fig:4} 
		\begin{center} 
			\includegraphics[width=12cm, height=9cm]{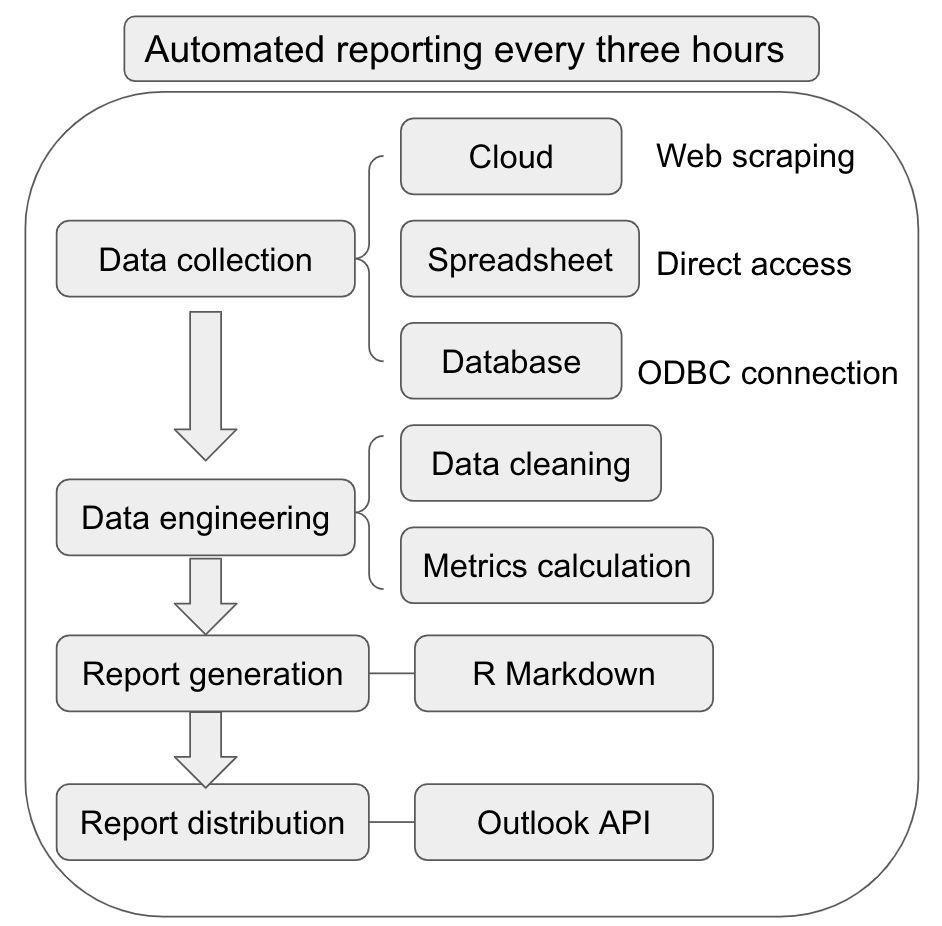}\hspace{0.1cm} 
			\caption{Overview of mission critical automation bot. Following a three-hourly schedule, the bot first collects data from difference sources such as cloud systems, spreadsheets and databases. After proper wrangling and reshaping of data, an operational status report is generated using R Markdown and distributed via Microsoft Outlook API.}
			\label{fig:bot} 
		\end{center} 
	\end{figure}

	With the right data available, the mission critical bot then performs extensive data engineering to properly massage the data, followed by report knitting using RMarkdown and report distribution by calling the Microsoft Outlook API. The whole automation process is able to generate and distribute the status report using around 5 minutes, and is scheduled to run every hour via a batch job in Windows Task Scheduler, as shown in Figure \ref{fig:bot}. 
	
	Besides automating reporting solution, the bot also provides recommendation on vehicle deactivation, saving operational cost if it is identified safe to do so. As mentioned earlier, the historical distributions of the number of jobs by job type, hour of the day and day of the week are also analyzed. By assessing the current volume with a certain percentile of the distribution on the historical hourly volume, operations can decide if one or more vehicle can be deactivated safely for the next few hours. For example, if the minimum hourly balance between supply and demand, based on the real time operational data, exceeds one vehicle for the next 12 hours, the bot will recommend standing down one disposal vehicle. Planners/controllers also have the flexibility to tell the bot which percentile they would like to choose when comparing current volume with historical distributions. A high percentile means a relatively conservative approach towards serving customers with high service level, but may risk having too many vehicles on stand by, while a low percentile indicates a relatively aggressive attitude towards serving customers with just enough vehicles, but may not reserve enough vehicles for sudden surges in demand. By constantly adjusting the percentile or threshold for different hours of the day and days of the week, the planners/controllers were able to improve the service level to nearly 100\% by minimizing the no-vehicle-available instances which used to happen on a daily basis, and save up to S\$3.2 million after adopting this bot as a key decision-making reference.

	\section{Implementation and Evaluation}
	
	To maximize the likelihood of adoption, the limo operational managers are also given the final right to edit, often to increase, the supply provision and generated schedule from the system, based on their operational experience, to deal with any last-minute changes not captured by the system. After all, they are the stakeholders who have eventually to bear the cost and answer to senior management. 
	
\begin{figure}[htbp]
		\begin{center} 
			\includegraphics[width=16cm, height=9cm]{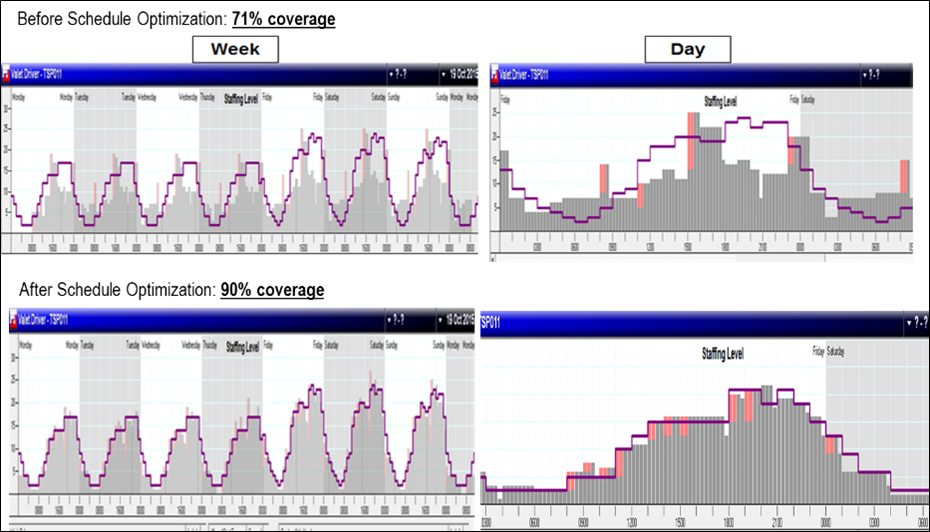}\hspace{0.1cm}
			\caption{Impact of the new system on demand coverage. More than 90\% of service requests are accommodated by optimized shift schedule, showing an improved coverage of nearly 20\% compared with previous experience-based scheduling approach.}
			\label{fig:outcome} 
		\end{center} 
	\end{figure}

	To better understand the system effectiveness and the need to do editing, the team created additional metrics in the daily management dashboard, to understand the number of cars originally scheduled by the system, the actual cars scheduled after editing (if any), and the actual number of cars in use. All three metrics were measured by hour of the day. Figure \ref{fig:outcome} shows the performance before and after the adoption of the system as shown in the dashboard. It was discovered that	scheduled shifts are enough to cover the actual number of cars needed for over 90\% of the cases, with no need for human edits. 
	
	The project was also rolled out in a gradual manner, introducing the system’s recommendations slowly and gradually, until the limo operational managers trusted the recommendations of the system with enough confidence. For example, in the initial days, an activation of 25 disposal vehicles was often observed across the busy evening hours on weekends, although the system recommended only 10. Instead of making a drastic cut to follow the system's recommendation, the managers are allowed to reduce the number first to 20 in the following weeks, and then to 15 and to 10 gradually, if no major complaints were observed. The success of the adoption of the system, in some sense, was partially due to project team's ability to convince the managers that the oversized schedule in the old approach had led to a big buffer compared with the new system’s recommendation. The reduction in outsourced labor due to the system's recommendation was also welcomed by most in-house drivers, since they have more opportunities to interact with guests and to earn tips. The reduced reliance on disposal vehicles also pushed the vendor towards more lean operations, such as engaging the same contract driver across consecutive shifts to reduce switching cost among different drivers.
	
\begin{figure}[htbp]
		\begin{center} 
			\includegraphics[width=12cm, height=8cm]{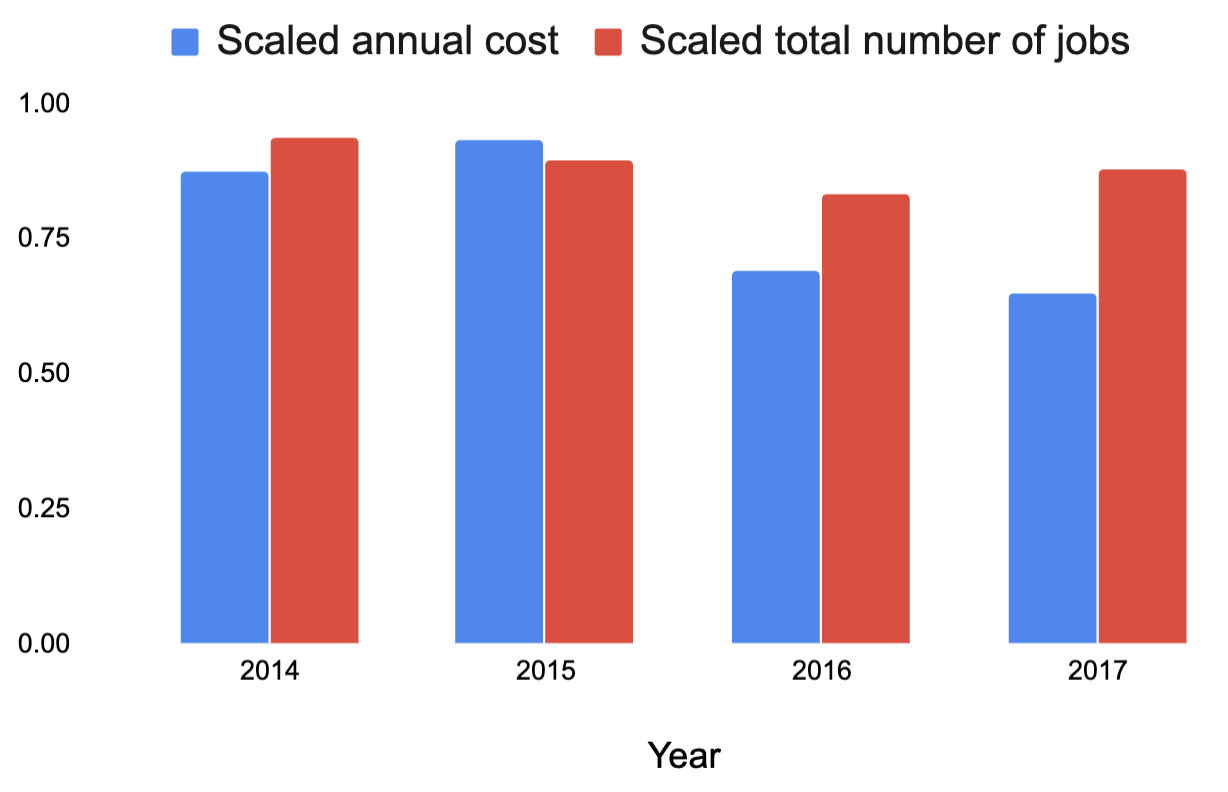}\hspace{0.1cm}
			\includegraphics[width=12cm, height=7cm]{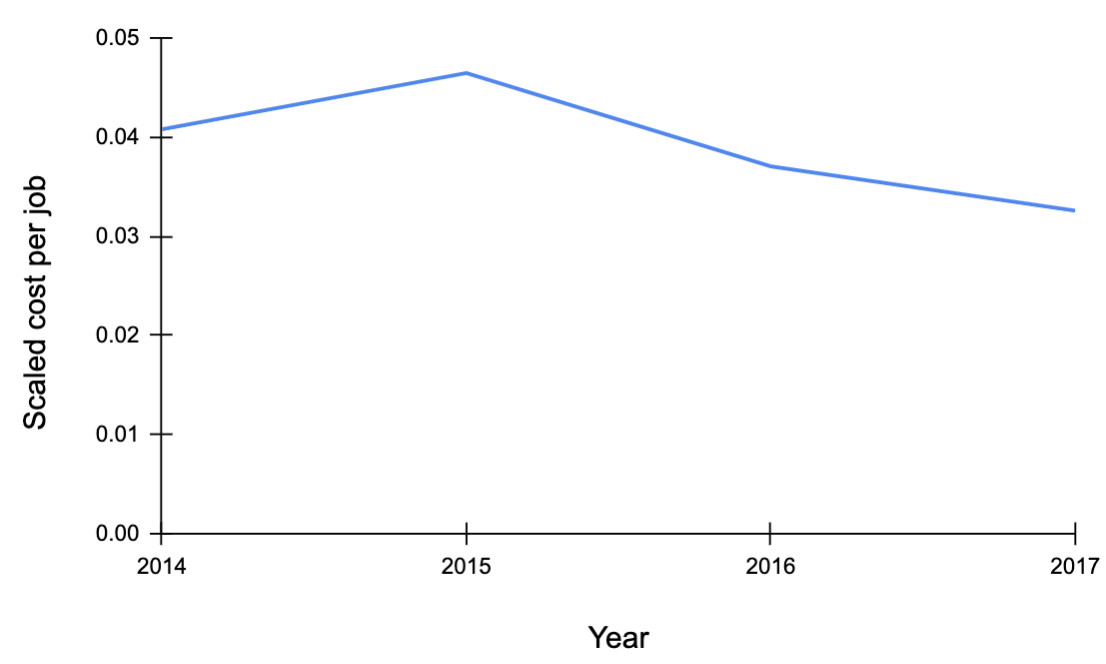}\hspace{0.1cm} 
			\caption{Plot of scaled total annual cost and number of jobs per year (top), and scaled cost per job (bottom). A relatively big decrease in annual cost is observed in 2016, the year when the system was implemented. Although the cost saving is partly accounted for by lower business volume, a further reduction in cost despite of increasing volume in the following year further demonstrates the effectiveness of the new system. This is also illustrated by the metric of total cost per job in the second figure, which is a more robust measure on cost reduction in view of change in business volume.}
			\label{fig:cost} 
		\end{center} 
	\end{figure}

	A significant impact from the new system on service level and cost savings is observed: up to a 99.98\% service level was achieved for over 100,000 jobs per year in comparison with frequent no-vehicle-available instances in the past; Net Promoter Score increased from 89.1\% to 92\% year-on-year in October, showing an improvement of 2.9\% since adoption of the new approach; a total cost saving of S\$ 3.2 million in the first year and more in following years, as illustrated in Figure \ref{fig:cost}.  Besides reduction in total cost, another effective and consistent measure is cost per job. As shown in Table \ref{T:annual_cost_per_job}, a substantial decrease in cost per job is observed since adoption of the forecasting and optimization system. After accounting for other factors such as depreciation of in-house vehicles, the cost per job decreased by over 20\% between 2015 and 2016 while total volume decreased by 7.5\%, with a further reduction of 12\% in cost per job between 2016 and 2017 despite a 6.3\% increase in total volume. 
	
	\begin{table}[htbp]
		\resizebox{\textwidth}{!}{%
			\begin{tabular}{ccccccccc}
				\multirow{2}{*}{} &
				\multicolumn{2}{c}{2014} &
				\multicolumn{2}{c}{2015} &
				\multicolumn{2}{c}{2016} &
				\multicolumn{2}{c}{2017} \\
				& {Total \$} & {\$/Job} & {Total \$} & {\$/Job} & {Total \$} & {\$/Job} &  {Total \$} & {\$/Job} \\
				Total Fixed Cost & 1,000 & 0.02147 & 1,209.98 & 0.02263 & 1,672.07 & 0.02205 & 1,794.52 & 0.02147 \\
				Depreciation & 503.36 & 0.01081 & 673.29 & 0.01259 & 571.72 & 0.00767 & 548.62 & 0.00657 \\
				Total Variable Costs & 262.37 & 	0.00563	 & 238.76 & 	0.0045	 & 291.07	 & 0.00384	 & 330.99 & 	0.00396 \\
				Total In-house Cost	 & 1765.73 & 	0.03791 & 	2122.02 & 	0.03968 & 	2544.86 & 	0.03357 & 	2674.13 & 	0.032 \\
				Total Outsourced Cost & 	3083.36 & 	0.04262 & 	3145.03 & 	0.05254 & 	1338.1 & 	0.04619 & 	951.72 & 	0.03435 \\
				Total Limo Cost (In-house and Outsourced)	 & 4849.09 & 	0.0408 & 	5267.05 & 	0.04647 & 	3882.96 & 	0.03706 & 	3629.7 & 	0.03259 \\

			\end{tabular}
			
		}
		\caption{Breakdown of annual total cost and cost per job scaled by a common denominator of S\$1,000 total fixed cost in 2014}
		\label{T:annual_cost_per_job}
	\end{table}


	

	The implementation of this system in the hotel has been a resounding success. The following are the summary of the documented benefits derived from this project:
	
	
	\begin{itemize}
		\item Achieved 99.97\% availability of limousine cars for patrons, a vast improvement compared to previous years. Only 22 instances of no vehicle available with 93,721 jobs completed from Jan-Nov 2016 compared to a few instances on daily basis in 2015
		\item  Increase in internal fleet productivity resulted in reduced dependency on outsourcing services. This is highlighted by a savings of S\$3.2 million in limo expense for 2016 compared against internal budgeted amount, a lowered cost per job by 21\% in 2016 compared to 2015, and a reduced commitment of the permanent outsourced base fleet in the limousine fleet from 10 to 9 in 2017.
		
		\item Enhanced Limo billing and verification process which drastically reduced billing discrepancies, and ensured timely payment to vendor. Using the new integrated system, the operations team identified and recovered overcharges equivalent to 5\% of total outsourced limo expense for 2015. 
	\end{itemize}
	
	The above automation and tools also allowed the team to brainstorm and identify further potential areas of improvement,  such as spotting certain trends in service delivery, that will bring about a better experience for the hotel's guests. For instance, vehicles travelling between the airport and the hotel typically require the same amount of time. Upon analyzing the data from the system's integrated dashboard, the team noticed that certain drivers would take longer to return from their jobs. This led them to identify a way to resolve the issue of staff accountability – which is to install a GPS tracker on every vehicle. This change not only allowed the team to track the vehicle’s location in real time, but also provided information such as speed of the vehicle. Since the implementation, a higher level of staff engagement and more consistent work performance are observed. The data also enabled management to make more effective decisions pertaining to scheduling, timing of breaks, and job allocations. 
	
	
	Moreover, the system developed for this problem is also used to manage the productivity and performance of the department, presenting an individualized 360-degree performance evaluation dashboard for each in-house driver. For example, key metrics such as the number of jobs completed and the average job duration for each individual driver are visualized together with the distribution of these metrics across the whole fleet, so as to indicate the relative performance ranking of the driver. Such automated performance dashboard also facilitates better team member engagement on a regular basis. 
	

	
	\section{Concluding Remarks}\label{sec:Result} 
	This paper explained how the time series demand forecasting methods, mission critical scheduling optimization and operational dashboards are applied to drive significant cost savings and a big increase in service level, for a major hotel in Singapore. Experiences in using open-source technologies to alleviate a staff shortage problem that was impeding business growth are also discussed. In terms of business benefits, the proposed approach was able to help operations achieve service level close to 100\% while maintaining a fixed fleet size without having additional permanent planners/controllers or a major increase in outsourcing, thus greatly improving resource utilization.

	\section{Appendix}\label{sec:appendix} 
	\subsection{Details on TBATS Model}
	The TBATS model consists of the following elements.
	
	
	\begin{eqnarray}
	y_t^{(\omega)} = \begin{cases} 
	\frac{y_t^{(\omega)}-1}{\omega} \quad \omega \neq 0 \\ 
	\log y_t \quad \omega=0 
	\end{cases} \\ 
	y_t^{(\omega)} = l_{t-1}+\phi b_{t-1}+\sum^T_{i=1}s^{(i)}_{t-m_i}+d_t\\ 
	l_t = l_{t-1}+\phi b_{t-1}+\alpha d_t\\ 
	b_t = (1-\phi)b+\phi b_{t-1}+\beta d_t\\ 
	s_t^{(i)} = s_{t-m_i}^{(i)}+\gamma_id_t\\ 
	d_t = \sum^p_{i=1}\varphi_id_{t-i}+\sum^q_{i=1}\theta_i\epsilon_{t-i}+\epsilon_t 
	\end{eqnarray}

	\noindent Here, $y_t^{(\omega)}$ is the inflated hourly volume at time $t$ with parameter $\omega$ used for Box-Cox transformation, $m_1, ..., m_T$ denote seasonal periods, $l_t$ is the local level in period $t$, $b$ is the long-run trend, $b_t$ is the short-run trend in $t$, $s_t^{(i)}$ represents the $i^{th}$ seasonal component at time $t$, $d_t$ denotes an ARMA($p$,$q$) process and $\epsilon_t$ is a Gaussian white noise process with zero mean and constant variance $\sigma^2$. The smoothing parameters are given by $\alpha$, $\beta$ and $\gamma_i$ for $i=1,...,T$. The damped trend has a damping parameter $\phi$, which is supplemented by a long-run trend $b$. The arguments for the model are $(\omega,\phi,p,q,m_1,...,m_T)$, making for a large number of parameters to be estimated. In order to have a more flexible and parsimonious approach, \cite{forecast2010} introduced the following trigonometric representation of seasonal components based on Fourier series: 
	
	\begin{eqnarray}
	s_t^{(i)} = \sum^{k_i}_{j=1}s^{(i)}_{j,t}\\ s^{(i)}_{j,t} = s^{(i)}_{j,t-1}cos\lambda_j^{(i)}+s^{*(i)}_{j,t-1}sin\lambda_j^{(i)}+\gamma_1^{(i)}d_t\\ 
	s^{*(i)}_{j,t} = -s_{j,t-1}sin\lambda_j^{(i)}+s^{*(i)}_{j,t-1}cos\lambda_j^{(i)}+\gamma_2^{(i)}d_t  
	\end{eqnarray}
	
	\noindent Note that $\gamma^{(i)}_1$ and $\gamma^{(i)}_2$ are smoothing parameters and $\lambda^{(i)}_j=2\pi j/m_i$. Here, $s^{(i)}_{j,t}$ represents the stochastic level of the $i^{th}$ seasonal component, while $s^{*(i)}_{j,t}$ represents the stochastic growth in the level of the $i^{th}$ seasonal component, i.e. change in the seasonal component over time. The number of harmonics required for the $i^{th}$ seasonal component is denoted by $k_i$. Most seasonal components will likely require fewer harmonics, thus reducing the number of parameters to be estimated from that of the previously described approach. Therefore, the seasonal component $s^{(i)}_t$ in equation (5) is replaced by trigonometric seasonal formulation, and equation (2) becomes $y^{(w)}_t=l_{t-1}+\phi b_{t-1}+\sum^T_{i=1}s^{(i)}_{t-1}+d_t$.

	\subsection{Details on Fleet Scheduling via Mathematical Programming}
	
	The scheduling problem can formulated as follows. 
	\begin{eqnarray*} 
		\mbox{minimize}\ &\sum_{j\in {\cal S}}l(j)x_{j}& \\ 
		s.t. \ &\sum_{j: s(j)\leq t < e(j)}x_j &\geq b(t), \ \ t=1,...,24 \\ 
		&\ \ \ \ \sum_{j\in {\cal S}}y_{j}&\leq N \\ 
		& \ \ \ \ \ \ \ \ \ \ 0\le x_{j} &\le My_j, \ \ j\in {\cal S} \\ 
		&\ \ \ \ \ \ \ \ \ \ \ \ \ \ y_{j} &\in \{0,1\} , \ \ j\in {\cal S} 
	\end{eqnarray*} 
	where $M$ denotes the maximum number of drivers deployed per shift and takes a value of 60. $b(t)$ is a for preset buffer to be covered at hour t, which takes on a value of zero or a positive integer, based on the forecast obtained. $N$ is the maximum number of disposal shifts allowed, and in this case, six.



	%
	%
	%

	
	\bibliographystyle{informs2014} 
	\bibliography{ref} 

\begin{thebibliography}{4}
\providecommand{\natexlab}[1]{#1}
\providecommand{\url}[1]{\texttt{#1}}
\providecommand{\urlprefix}{URL }

\bibitem[{Cheikhrouhou et~al.(2011)Cheikhrouhou, François, Omar,
  \protect\BIBand{} Philippe}]{combine2011}
Cheikhrouhou N, François M, Omar A, Philippe W (2011) A collaborative demand
  forecasting process with event-based fuzzy judgements. \emph{Computers \&
  Industrial Engineering} 61:409--421.

\bibitem[{Chun(2010)}]{limo2010}
Chun H (2010) Optimizing limousine service with ai. \emph{Innovative
  Applications of Artificial Intelligence Conference} 10.

\bibitem[{Livera et~al.(2011)Livera, Hyndman, \protect\BIBand{}
  Snyder}]{forecast2010}
Livera A, Hyndman R, Snyder RD (2011) Forecasting time series with complex
  seasonal patterns using exponential smoothing. \emph{Journal of the American
  Statistical Association} 106(496):1513--1527.

\bibitem[{Wan et~al.(2020)Wan, Ghazzai, \protect\BIBand{} Massoud}]{taxi2020}
Wan X, Ghazzai H, Massoud Y (2020) A generic data-driven recommendation system
  for large-scale regular and ride-hailing taxi services. \emph{Electronics}
  9(648).

\end{thebibliography}

	
\end{document}